\documentclass[%
 reprint,
 amsmath,amssymb,
 aps,
prb,
superscriptaddress,
longbibliography
]{revtex4-2}

\usepackage{color}
\usepackage{soul}
\usepackage[english]{babel}
\usepackage{comment}
\usepackage{graphicx}
\usepackage{dcolumn}
\usepackage{bm}
\usepackage{hyperref}
\usepackage{xr}

\begin{document}

\preprint{APS/123-QED}
 
\title{Towards Excitations and Dynamical Quantities in Correlated Lattices \\ with Density Matrix Embedding Theory}

\author{Shuoxue Li}
\email{sli7@caltech.edu}
\affiliation{Division of Chemistry and Chemical Engineering, California Institute of Technology, Pasadena, California 91125, USA}
\author{Chenghan Li}
\affiliation{Division of Chemistry and Chemical Engineering, California Institute of Technology, Pasadena, California 91125, USA}
\author{Huanchen Zhai}
\thanks{Present address : Initiative for Computational Catalysis, Flatiron Institute, 160 5th Avenue, New York, NY 10010, USA}
\affiliation{Division of Chemistry and Chemical Engineering, California Institute of Technology, Pasadena, California 91125, USA}
\author{Garnet Kin-Lic Chan}
\email{gkc1000@gmail.com}
\affiliation{Division of Chemistry and Chemical Engineering, California Institute of Technology, Pasadena, California 91125, USA}

\begin{abstract}
Density matrix embedding theory (DMET) provides a framework to describe ground-state expectation values in strongly correlated systems, but its extension to dynamical quantities is still an open problem. We show one route to obtaining excitations and dynamical spectral functions by using the techniques of DMET to approximate the matrix elements that arise in a single-mode inspired excitation ansatz. We demonstrate this approach in the 1D Hubbard model, comparing the neutral excitations, single-particle density of states, charge, and spin dynamical structure factors to benchmarks from the Bethe ansatz and density matrix renormalization group. Our work highlights the potential of these ideas in building computationally efficient approaches for dynamical quantities.
\end{abstract}

\maketitle

\section{\label{sec:level1} Introduction}

The study of excitations is central to understanding condensed matter systems. This is especially the case in strongly correlated materials, where the low-energy excitations can be used to diagnose the presence of interesting phenomena such as high-temperature superconductivity\cite{htsc-li,htsc-zhao,restrepoCoupledElectronicMagnetic2022}, quantum criticality\cite{Sachdev2011-xf,quantum-criticality-kang} and various types of magnetism\cite{magnetism-atanasov}. 

To describe the excitations of correlated materials at an affordable cost, here we attempt to develop a method based on density matrix embedding theory (DMET)\cite{dmet-original-knizia,dmet-practical-guide-wouters}, which was originally introduced to approximate ground-state properties of correlated fermionic systems\cite{Zheng2016-ie,dmet-abinitio-cui,Cui2022-tv,cuiInitioQuantumManybody2025}.
DMET is a quantum embedding method~\cite{Jones2020-kk,Sun2016-ch}, which approximates expectation values in the bulk ground-state from a set of small quantum many-body problems involving impurities (or fragments) of the bulk problem.
 DMET has also been applied to localized and impurity-like excitations ~\cite{dmet-ext-colorcenter-mitra,Verma2023-bp,Lau2024-ke,dmet-ext-smm-ai}, and there have been attempts to extend the framework to  delocalized and collective excitation spectra in bulk systems~\cite{Booth2015-bl,Sriluckshmy2021-dz,Scott2021-kf}.
 This work is focused on the latter problem.

Spectral DMET~\cite{Booth2015-bl} and energy weighted DMET~\cite{Sriluckshmy2021-dz,Scott2021-kf} are two prior attempts in the literature to extend DMET to treat delocalized excitations. These can be characterized as improving the description of excitations by generalizing the definition of the impurity, for example through the frequency-dependent bath in spectral DMET, or by extending the bath to capture Hamiltonian moments in EwDMET, however, neither approach provides direct access to the delocalized many-particle excited states. For example, spectral DMET only provides access to a (correlated) description of the impurity dynamical correlation functions, while EwDMET only provides access to the bulk single-particle Green's function. Here we explore an alternative approach that allows us to approximate the collective wavefunctions of the excitations, thereby also allowing access to all bulk dynamical correlation functions. To do so we introduce a basis of local excitation operators on top of the ground-state to span the relevant space of low-energy excitations. This is similar to the tangent-space idea in tensor networks~\cite{verstraeteValencebondStatesQuantum2004,peps-verstraete,peps-fermion-corboz,tangent-space-abinitio-nakatani,tangent-space-haegeman,tangent-space-tutorial-vanderstraeten} and is a variant of the well-known single mode approximation~\cite{Feynman1953-ns,Girvin1985-as,Repellin2014-xz}. The projection of the Hamiltonian into this space results in an eigenvalue problem for a subset of the local excitations. The ideas of DMET then enter into the approximation of the matrix elements of this Hamiltonian in the space of excitations. These generalize the approach taken in DMET to approximate ground-state operator expectation values to expectation values between different states. This allows us to describe the collective excited state of the whole lattice by explicitly considering the correlation between sites.

We first recap the theory of ground-state DMET in section~\ref{sec:method} before introducing the excitation ansatz and the DMET prescription for approximating matrix elements in the excitation basis. 
In section \ref{sec:results}, we then assess this approach in the computation of excitation energies and spectra of the 1D Hubbard model, benchmarking against Bethe Ansatz and density matrix renormalization group results.  We conclude with analysis of the potential and pitfalls of our framework. 

\section{\label{sec:method} Methods}

\subsection{\label{subsec:dmet-gs}Ground-state Density Matrix Embedding Theory and its Extension to Excited States}

We first recap some essentials of the ground-state DMET formalism. We consider partitioning a system into a set of sites that we term an  impurity (or fragment) $I$  with the remainder denoted as the environment. Then the ground-state wave-function of the system Hamiltonian $\hat{H}$ can be decomposed between the impurity and environment Hilbert spaces as
$|\Psi_0^I\rangle = \sum\limits_m  \psi_m |  \alpha_m \rangle \otimes |  \beta_m \rangle$, where we say that  $\{|{\alpha}_m\rangle \otimes |{\beta}_m\rangle\}$ defines an embedding Hilbert space for the ground-state, i.e. the ground-state of $\hat{H}$ projected into the embedding space coincides with the ground-state of $\hat{H}$. The main ideas are then to (i) approximate the embedding space of $\Psi_0^I$ via the embedding space of an auxiliary mean-field ground-state $\Phi_0^I$ (that can be improved self-consistently), and (ii) to approximate expectation values of $\Psi_0$ by assembling expectation values of approximate $\Psi_0^I$ computed for partitionings of the system into different impurities $I$ and their environments. The detailed ground-state DMET algorithm is discussed in many places, see Ref. \cite{dmet-practical-guide-wouters}.

Assume that we have computed a set of $\Psi_0^I$ for different impurities of the system from the above procedure. Our work relies on extending approximation (ii), which we now illustrate. Say we wish to approximate an element of the single-particle density matrix. Then we use the formula called "democratic partitioning" in Ref. \cite{dmet-practical-guide-wouters}:
    \begin{equation}
        \langle \Psi_0 | \hat a^\dagger_i \hat a_j | \Psi_0 \rangle \approx \dfrac{1}{2} ( \langle \Psi_0^{i} | \hat a^\dagger_i \hat a_j | \Psi_0^{i} \rangle + \langle \Psi_0^{j} | \hat a^\dagger_i \hat a_j | \Psi_0^{j} \rangle )
        \label{eq:democratic-partitioning}
    \end{equation}
    in which $| \Psi_0^{i} \rangle$ and $| \Psi_0^{j} \rangle$ are DMET ground-states for partitionings where sites $i$ and sites $j$ are included in the impurities, respectively. In this way, expectations of non-local operators can be evaluated by assembling small ground-state DMET calculations involving different choices of impurities.

    Our strategy to approximate excited states in DMET will involve building an excitation basis by applying operators to the ground-state and solving for the excited states in the corresponding linear space, with matrix elements approximated by the above procedure. We describe each of these steps in further detail below.

\subsection{\label{subsec:ts}Excitation basis}

\subsubsection{Single-site projector excitation basis} In order to construct the low-energy excitation basis for the correlated system, 
we define a set of operators that act on the ground-state wavefunction. 

Specifically, in the simplest case, we consider:
\begin{equation}
    | \Psi_{s_i} \rangle = \hat P_{s_i} | \Psi_0 \rangle
    \label{eq:ts-project}
\end{equation}
where $| \Psi_0 \rangle$ is the ground-state wave function and $\hat P_{s_i}$ is a projector onto the states of site $i$, i.e. $\hat{P}_{s_i} = |s_i \rangle \langle s_i|$.

In second quantization, for a single site, we can choose them to be \cite{bunemannMultibandGutzwillerWave1998}
\begin{equation}
    \begin{aligned}
    \hat P_{\mathrm{vac}} &= (1-\hat n_\alpha)(1- \hat n_\beta)\\
    \hat P_\uparrow &= \hat n_\alpha (1-\hat n_\beta) \\
    \hat P_\downarrow &= (1-\hat n_\alpha)\hat n_\beta \\
    \hat P_{\uparrow\downarrow} &= \hat n_\alpha \hat n_\beta
    \end{aligned}
    \label{eq:definition-projector}
\end{equation}
where $\hat n_\alpha$ and $\hat n_\beta$ are density operators for spin $\alpha$ and $\beta$ electrons on site $i$: $\hat n_\sigma=\hat a^\dagger_{\sigma} \hat a_\sigma,\, \sigma = \alpha, \beta$. 
The excited state ansatz is then
\begin{equation}
    |\Psi\rangle = \sum_i \sum_{s_i} c_{i, s_i} |\Psi_{s_i}\rangle = \sum_i \sum_{s_i} c_{i, s_i} \hat{P}_{s_i} |\Psi_0\rangle \label{eq:smansatz}
\end{equation}
and the excitation energy and coefficients can be obtained by diagonalizing the effective eigenvalue problem
\begin{equation}
     \mathbf{H C = S C \omega}
     \label{eq:heff-eigenvalue-problem}
 \end{equation}
 in which the elements of effective Hamiltonian matrix $\mathbf H$ and overlap matrix $\mathbf S$ are

\begin{equation}
    \begin{aligned}
         H_{s_i, s_j} &= \dfrac{1}{2} \langle \Psi_0 |  [\hat P_{s_i}^\dagger, \hat H] \hat P_{s_j} + \hat P_{s_i}^\dagger [\hat H, \hat P_{s_j}] | \Psi_0 \rangle \\
          S_{s_i, s_j} &= \langle \Psi_{s_i} |\Psi_{s_j}\rangle 
    \end{aligned}
    \label{eq:commutator}
\end{equation}

where we have used the commutator formulation so that the eigenvalues that appear are the excitation energies (i.e. the difference between the excited state and ground state energies). 
In a translationally invariant system, this can be viewed as a single-mode equation-of-motion excitation ansatz \cite{Feynman1953-ns,Girvin1985-as,Repellin2014-xz}, which is also commonly used in tensor network treatments of excitations in the tangent space \cite{non-redundant-ts-wouters,tangent-space-abinitio-nakatani,tangent-space-haegeman,tangent-space-tutorial-vanderstraeten}.

\subsubsection{Generalized excitation basis} As a generalization of the above excitation operator ansatz involving single sites, we can consider excitation operators for a set of sites (which we refer to as an excitation patch)
\begin{equation}
    I \equiv \{i, j, k, \cdots\}
\end{equation}
In addition, we can extend the operators from projectors to generalized operators, e.g. of the form $\hat{o}_{s_i, s_i'} = |s_i \rangle \langle s_i'|$ for a single site. We refer to the corresponding excitation basis as the generalized excitation basis.
A complete set of single-site operators $\hat o_{s_i, s_i'}$ is shown in Table \ref{table:operators}. 
Within the single-mode like ansatz used in this work,
the excitation operators of a patch $\hat O_{s_I, s_I'} = \hat o_{s_i, s_i'} \hat o_{s_j, s_j'} \hat o_{s_k, s_k'} \cdots $ should be chosen to be consistent with the quantum numbers of the excitation to be studied. For example, for a neutral excitation, $\hat O_{s_I, s_I'}$ should be restricted to conserve the particle number.

\begin{table}[htbp]
    \centering
    \begin{tabular}{ccc}
    \hline
        Single-site operators & $N$ & $S_z$ \\ \hline
       $\hat n_\alpha \hat n_\beta, \hat n_\alpha \hat m_\beta, \hat m_\alpha \hat n_\beta, \hat m_\alpha \hat m_\beta$ & 0 & 0 \\ 
        $\hat a^\dagger_\alpha \hat n_\beta, \hat a^\dagger_\alpha \hat m_\beta$ & 1 & 1/2 \\ 
        $\hat a_\alpha \hat n_\beta, \hat a_\alpha \hat m_\beta$ & -1 & -1/2 \\
        $\hat a^\dagger_\beta \hat n_\alpha,\hat a^\dagger_\beta \hat m_\alpha $ & 1 & -1/2 \\ 
        $\hat a_\beta \hat n_\alpha, \hat a_\beta \hat m_\alpha $ & -1 & 1/2 \\ 
        $\hat s_+ = \hat a^\dagger_\alpha \hat a_\beta$ & 0 & 1 \\ 
        $\hat s_- = \hat a^\dagger_\beta \hat a_\alpha$ & 0 & -1 \\
        $\hat a^\dagger_\alpha  \hat a^\dagger_\beta$ & 2 & 0 \\ 
        $\hat a_\alpha \hat a_\beta$ & -2 & 0 \\ \hline
    \end{tabular}
    \caption{Independent general single-site excitation operators and associated quantum numbers, where $\hat n_\alpha = \hat a^\dagger_\alpha \hat a_\alpha, \hat n_\beta = \hat a^\dagger_\beta \hat a_\beta, \hat m_\alpha = 1 - \hat n_\alpha, \hat m_\beta = 1 - \hat n_\beta.$}
    \label{table:operators}
\end{table}

\begin{figure}
\includegraphics[width=0.44\textwidth]{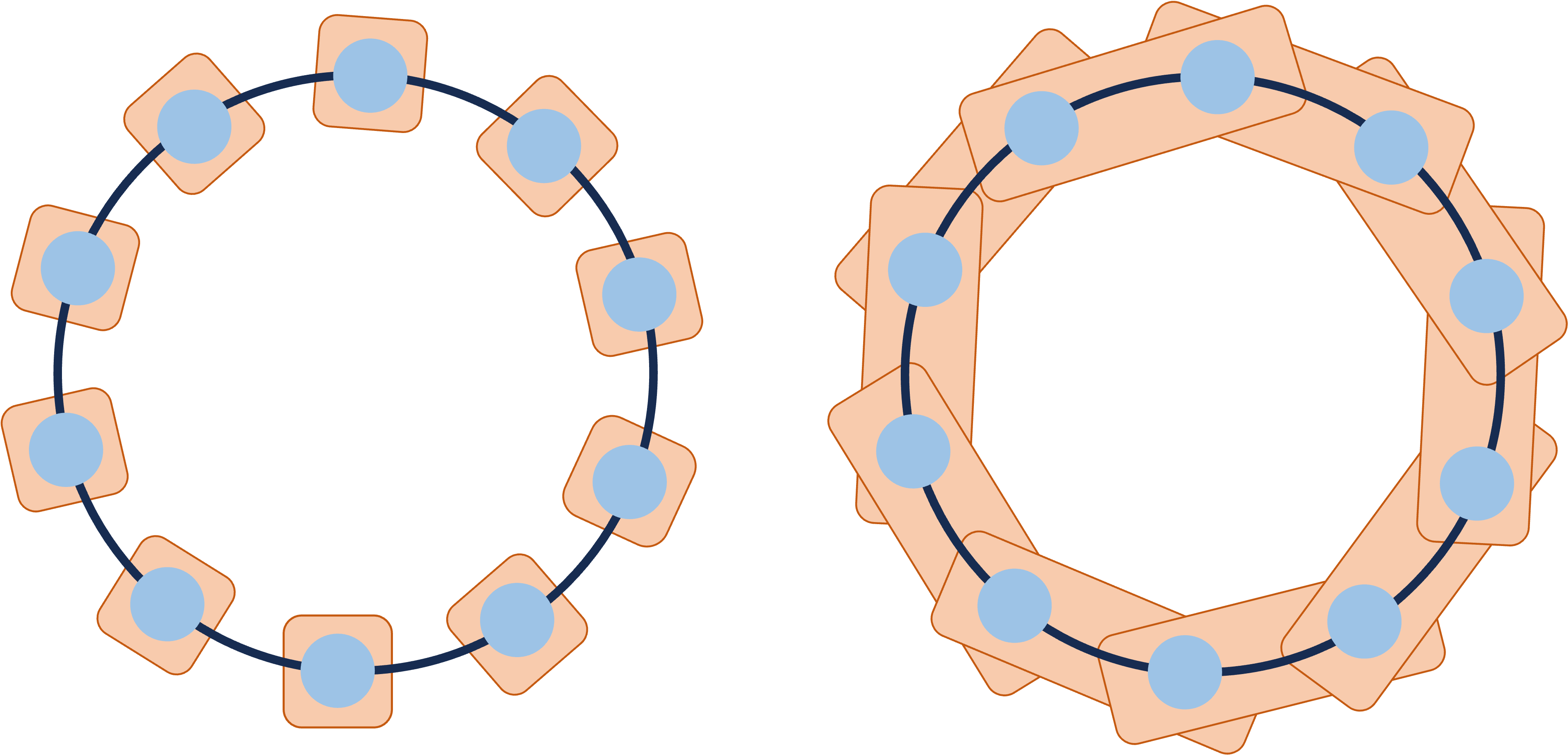}
\caption{\label{fig:tangent-space-choice} Illustration of single-site and double-site patches (red regions) for a 1D system with periodic boundaries.}
\end{figure}

For an $N$-site patch $I$, the total number of general excitation operators grows rapidly with the patch size. For instance, the total number of the neutral excitation operators $\hat O_{s_I, s_I'}$ is $(C_{2N}^N)^2$. Consequently, physically motivated truncations of the operator set are needed for large $N$. We will describe the use of both full and truncated operator bases associated with the patches in the calculations below. 

For systems with periodic boundaries, it is natural to choose identical, translated, patches to recover the global translational symmetry. For a 1D system, we illustrate our choices for one- and two-site patches in Fig.~\ref{fig:tangent-space-choice}. We then use momentum labels to reduce the number of coefficients in Eq.~\ref{eq:smansatz} and to block-diagonalize the effective Hamiltonian and overlap matrices. For example, for the one-site patch, the state with total lattice momentum $\pmb k$ is represented as
\begin{equation}
    |\Psi(\pmb k)\rangle = \sum_j \sum_{s} c_{s}(\pmb k) e^{-i\pmb  k \cdot \pmb R_j} |\Psi_{s_j}\rangle
\end{equation}

\subsection{DMET approximation of the effective Hamiltonian and overlap}

\label{sec:dmetapproximation}
\begin{figure}[htbp]
    \centering
    \includegraphics[width=0.35\textwidth]{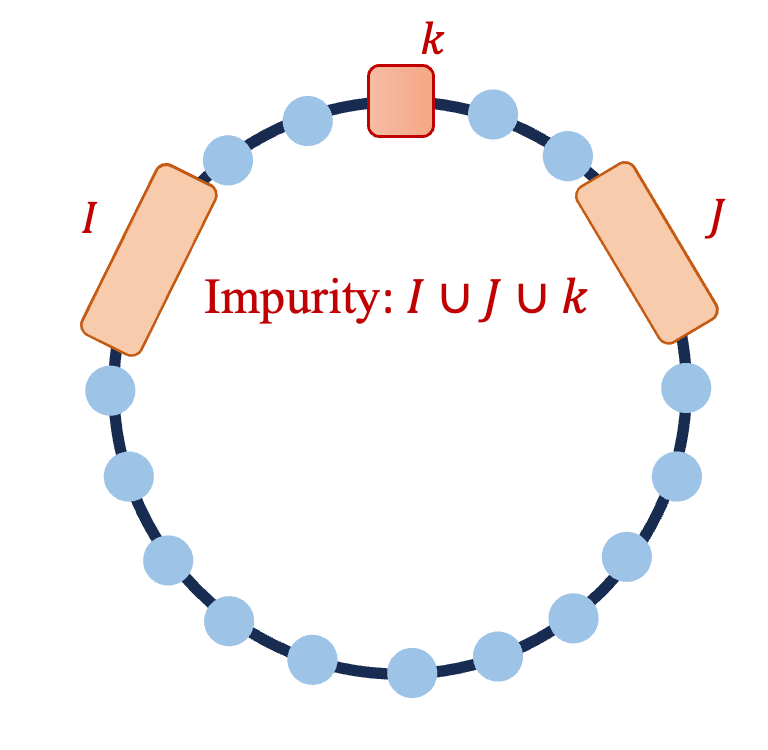}
    \caption{Illustration of a large impurity with the bra patch, ket patch and democratic partitioning index.}
    \label{fig:imp-illustration}
\end{figure}

Now we use DMET ideas to estimate the effective Hamiltonian and the overlap matrix elements in subsection \ref{subsec:ts} with DMET. (We limit ourselves below to examples for fermionic Hamiltonians with at most quartic terms). Expanding the commutator in Eq.~\ref{eq:commutator}, we find terms such as $\langle \Psi_0 |\hat O_{I}^\dagger \hat O_{J} \hat{H} |\Psi_0\rangle$ and
$\langle \Psi_0 |\hat O_{I}^\dagger \hat{H} \hat O_{J} |\Psi_0\rangle$ (here $\hat O_I$ can be taken to either be a projection operator $\hat O_{s_I, s_I}$or a generalized excitation operator $\hat O_{s_I, s_I'}$). The action of $\hat{O}_{I}$ and $\hat{O}_{J}$ on $\Psi_0$ can be computed exactly by using an impurity that contains the sites in $I \cup J$. Then, we consider the individual fermion terms in $\hat{H}$, and use the DMET democratic partitioning recipe \cite{dmet-practical-guide-wouters} to estimate the corresponding expectation value. For example, for $\hat{a}_k^\dagger \hat{a}_l$, we use the formula,

\begin{widetext}
\begin{equation}
        \langle \Psi_0 | \hat{O}_{I}^\dagger \hat a^\dagger_k \hat a_l \hat{O}_{J} | \Psi_0\rangle \approx 
         \dfrac{1}{2} ( \langle \Psi_{0}^{I\cup J \cup \{k\}} | \hat{O}_{I}^\dagger \hat a^\dagger_k \hat a_l \hat{O}_{J} | \Psi_{0}^{I\cup J \cup \{k\}} \rangle + \langle \Psi_{0}^{I\cup J \cup \{l\}} | \hat{O}_{I}^\dagger \hat a^\dagger_k \hat a_l \hat{O}_{J} | \Psi_{0}^{I\cup J \cup \{l\}} \rangle )
         \label{eq:trans-1rdm}
\end{equation}
\end{widetext}
where on the r.h.s. we have used the DMET democratic partitioning of expectation values in Eq. \ref{eq:democratic-partitioning} between impurity ground states, where the impurity includes one of the site indices of the operator being evaluated ($k$ or $l$ above) as well as the sites of $I$ and $J$, as demonstrated in Fig. \ref{fig:imp-illustration}.

Similarly, a matrix element like 
$ \langle \Psi_0 | \hat{O}_I^\dag  \hat a^\dagger_k \hat a^\dagger_l \hat a_m \hat a_n  \hat{O}_J |\Psi_{0} \rangle$ 
can also be approximated using the DMET democratic-partitioning method using impurities that include $I, J$ and one of the indices out of $(k, l, m, n)$, 
\begin{widetext}
\begin{equation}
    \begin{aligned}
        \langle \Psi_{I} | \hat a^\dagger_k \hat a^\dagger_l \hat a_m \hat a_n | \Psi_{J} \rangle \approx \dfrac{1}{4} (
        \langle \Psi_{I}^{I \cup J \cup \{k\}} | \hat a^\dagger_k \hat a^\dagger_l \hat a_m \hat a_n | \Psi_{J}^{I \cup J \cup \{k\}} \rangle+ \langle \Psi_{I}^{I \cup J \cup \{l\}} | \hat a^\dagger_k \hat a^\dagger_l \hat a_m \hat a_n | \Psi_{J}^{I \cup J \cup \{l\}} \rangle \\ +
        \langle \Psi_{I}^{I \cup J \cup \{m\}} | \hat a^\dagger_k \hat a^\dagger_l \hat a_m \hat a_n | \Psi_{J}^{I \cup J \cup \{m\}} \rangle + \langle \Psi_{I}^{I \cup J \cup \{n\}} | \hat a^\dagger_k \hat a^\dagger_l \hat a_m \hat a_n | \Psi_{J}^{I \cup J \cup \{n\}} \rangle)
    \end{aligned}
    \label{eq:trans-2rdm}
\end{equation}
\end{widetext}
where we have used the simpified notation $|\Psi_{I}\rangle = \hat{O}_{I} |\Psi_0\rangle$.
Using the above rules, it is possible to approximate all matrix elements  encountered in this work. Further details are given in Appendix \ref{append:democratic-partitioning}. 

For the overlap matrix $\langle \Psi_{I} | \Psi_{J} \rangle$, we analogously obtain

\begin{equation}
    \langle \Psi_{I} | \Psi_{J} \rangle \approx \langle \Psi_{I}^{I \cup J} | \Psi_{J}^{I \cup J} \rangle
    \label{eq:ovlp-eq}
\end{equation}

Because of the use of 3 patches $I, J, k$ that loop over the lattice in the above expressions, the computational scaling of the above calculations is cubic in system size. For translationally invariant problems, this is reduced to quadratic scaling, since only the relative positioning of $I, J, k$ contributes to a unique matrix element. For example, $I$ can be fixed to start at the first site and matrix elements computed for $J, k$, and matrix elements for a translated $I$ (and the $J, k$ obtained by translation) have the same value.

\subsection{Overlap truncation}

The overlap matrix defined in Eq.~\ref{eq:commutator} must have positive eigenvalues. However, because it is computed approximately, it may violate this exact condition. Further, the excitation basis can contain redundancies. For example, in Eq.~\ref{eq:definition-projector}, we have $\sum_{s_i} \hat{P}_{s_i}=1$, and these identity operators on each site lead to redundacies in the excitation basis. Similarly, for 
overlapping patches, operators that act only on the overlapping sites will be redundant between the different patches. Beyond these exact redundancies, the non-orthogonality of the excitation basis states may further lead to a large condition number in the overlap matrix. 

To handle these numerical issues, we remove some eigenvectors of the overlap matrix and solve for excitations in the remaining linear space.

\begin{figure}
    \centering
    \includegraphics[width=0.9\linewidth]{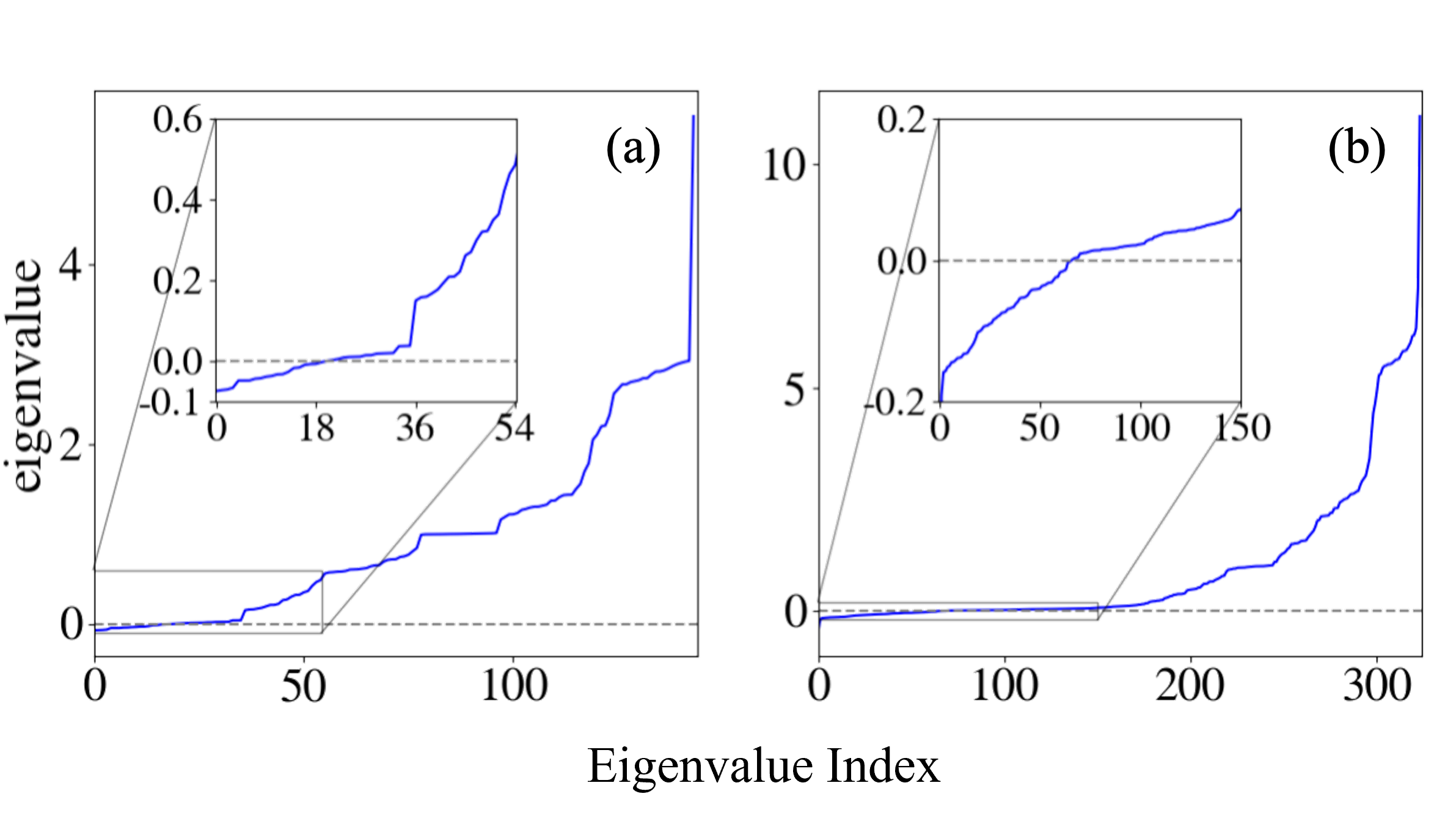}
    \caption{Sorted eigenvalues of the approximate overlap matrix of the excitation basis for a half-filled one-dimensional Hubbard model with $18$ sites, $U = 8$, from $k=\pi / 9$ to $k = \pi$ using two-site patches. (a) Excitation basis is the  projector basis; (b) Excitation basis is the generalized excitation basis.}
    \label{fig:overlap-spectra}
\end{figure}

In Fig. \ref{fig:overlap-spectra} which shows a test on the 18-site Hubbard model using two-site patches, negative eigenvalues of the approximate overlap matrix appear for both the projector excitation basis and the generalized excitation basis. However, we observe a clear jump in the overlap matrix eigenvalues at the 36th eigenstate when using the projector excitation basis, corresponding to the number of zero eigenstates in the exact overlap matrix (Fig. \ref{fig:overlap-spectra} (a) inset). When using  the generalized excitation ansatz (Fig. \ref{fig:overlap-spectra} (b)) we see a continuous change in the eigenvalues.  We thus use the following truncation strategy: (1) For the projector excitation basis, we discard a number of modes equivalent to the number of exact theoretical redundancies in the basis; (2) for all other cases, we set a cutoff and discard overlap eigenvectors with eigenvalues below this cutoff.  The influence of this cutoff is discussed in Section \ref{subsec:gf}.

\section{\label{sec:results}Results}

We now discuss some applications of the above formulation to the 1 band Hubbard model at half filling, described by the Hamiltonian
\begin{equation}
\begin{aligned}
 \hat H = \sum\limits_{i=1}^L \sum\limits_\sigma -t(\hat a^\dagger_{i, \sigma} \hat a_{i+1, \sigma} + \hat a^\dagger_{i+1, \sigma} \hat a_{i, \sigma} ) \\+ \sum\limits_{i=1}^L U \hat n_{i, \uparrow} \hat n_{i, \downarrow}
\end{aligned}
\label{eq:1d-hubbard-model}
\end{equation}
In all the calculations below, we start from a DMET ground-state without self-consistency, i.e. the auxiliary mean ground-state $\Phi_0^I$ is the ground-state of the $U=0$ Hamiltonian, which allows us to use the full translational symmetry of the problem, and we work in units with $t=1$.

\subsection{\label{subsec:energy-overview} Lowest excitation band}

We start with the approximation of the lowest band of excitation energies in the 1-band Hubbard model. To compute the expressions in Sec.~\ref{sec:dmetapproximation} 
we use full configuration interaction (FCI) and converged Density Matrix Renormalization Group (DMRG) calculations \cite{dmrg-original-white,dmrg-review-1-schollwock,dmrg-review-2-schollwock,dmrg-abinitio-theory-chan,zhaiBlock2ComprehensiveOpen2023b} to obtain the ground-state wavefunctions for the impurities, and to evaluate the matrix elements in Sec.~\ref{sec:dmetapproximation} . We then solve for the excited states using Eq.~\ref{eq:commutator} and compare to benchmark DMRG excitation energies on the same finite system, and from the Bethe Ansatz \cite{bethe_ansatz, hubbard-book-essler,hubbard-bethe-ansatz-lieb} (which yields the thermodynamic limit result). 

\begin{figure}
    \includegraphics[width=0.50\textwidth]{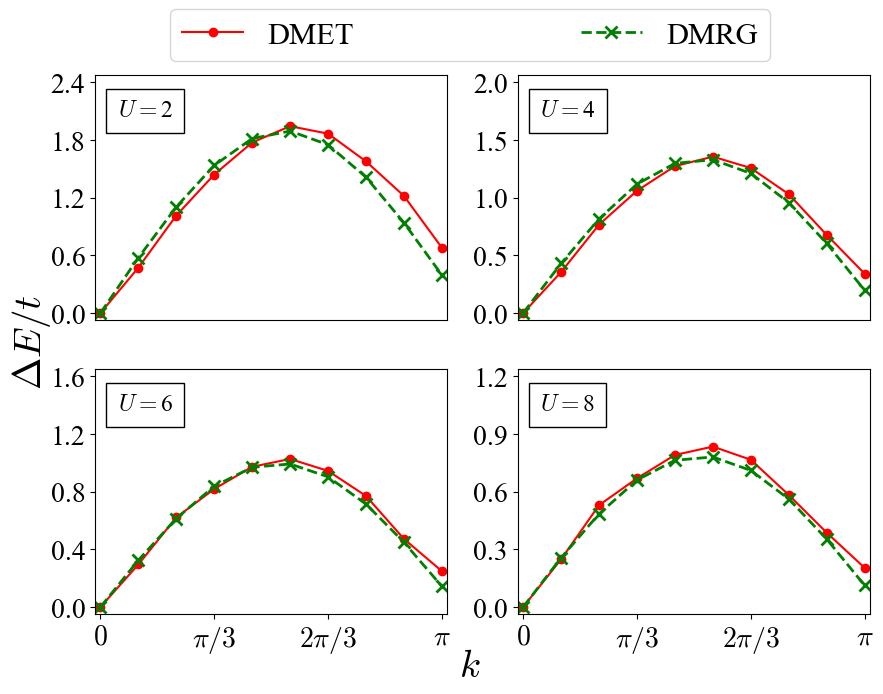}
    \caption{Comparison of DMET with DMRG excitation energies for the 1-band Hubbard model on 18 sites with periodic boundary conditions. We show the lowest excited state for each crystal momentum $k$. A 4-site patch is used to build an effective Hamiltonian for the excitations. For the DMRG calculation, bond dimension $M=1200$ is used to calculate the first 50 excited states with a discarded weight of $\sim 10^{-8}$.}
    \label{fig:energy-overview}
\end{figure}

Fig.~\ref{fig:energy-overview} shows the DMET and DMRG results for an 18-site Hubbard model with $U = 2, 4, 6$ and $8$ using a four-site patch, an excitation basis consisting of the projectors on each patch ($256$ projectors for each $k$ point, before eliminating redundancies). From the comparison to the DMRG benchmark, we find that the excitation basis and DMET approximation captures the energy momentum relationship accurately even with a small variational space.

\begin{figure}
    \centering
    \includegraphics[width=0.45\textwidth]{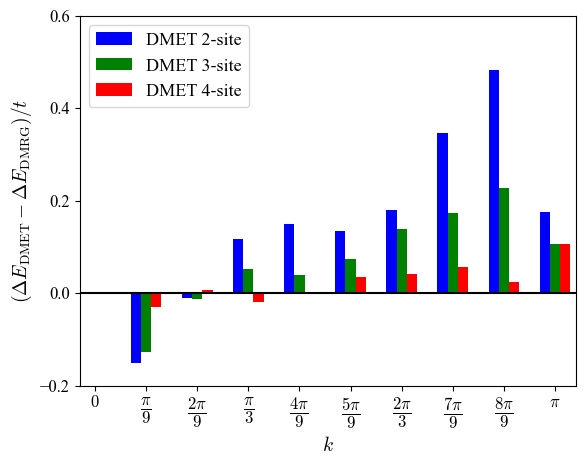}
    \caption{The difference betweeen the lowest excitation energy as a function of $k$ from DMET and DMRG for $L=18, U=6$. The DMRG results can be assumed to be exact on the scale of this plot, and we show the DMET results as a function of patch size.}
    \label{fig:energy-patchsize} 
\end{figure}

Fig.~\ref{fig:energy-patchsize} shows the performance of the DMET approximation with patch sizes 2, 3 and 4 for the excitation energies in the lowest excitation band. We see that the largest errors occur when $k \sim \pi$, but this error is significantly reduced when increasing the patch size.

\subsection{\label{subsec:gf} Single- and two-particle Green's functions}

We next compute one- and two-particle Green's functions using the excited states from 
the approximate Hamiltonian matrix. The general form of the Green's function we compute is

\begin{equation}
    G(\omega, \hat A, \hat B) = \langle \Psi_0 | \hat A^\dagger (\omega - (\hat H - E_0) + i\eta)^{-1} \hat B | \Psi_0 \rangle
\end{equation}
where $\hat A, \hat B$ are arbitrary operators.

We can compute $G$ from its Lehmann representation:
\begin{equation}
    G(\omega, \hat A, \hat B) = \sum\limits_{m} \dfrac{\langle \Psi_0 | \hat A^\dagger | \Psi_m \rangle \langle \Psi_m | \hat B | \Psi_0 \rangle}{\omega - (E - E_0) + i\eta}
\end{equation}
where $m$ is the electronic state index, and we compute the matrix elements as

\begin{equation}
    \begin{aligned}
        &\langle \Psi_0 | \hat A^\dagger | \Psi_m \rangle \approx \sum\limits_{I} c_m^{I} \langle \Psi_0 | \hat A^\dagger \hat O_{I} | \Psi_0 \rangle \\ &\langle \Psi_0 | \hat A^\dagger \hat O_{I} | \Psi_0 \rangle \approx \langle \Psi_0^{A \cup I} | \hat A^\dagger \hat O_{I} | \Psi_0^{A \cup I} \rangle
    \end{aligned}
\end{equation}
where 

$| \Psi_0^{A \cup I} \rangle$ denotes the DMET ground state where the impurity contains sites of operator $\hat A$ and patch $I$.

\begin{figure}[htbp]
    \centering
    \includegraphics[width=0.50\textwidth]{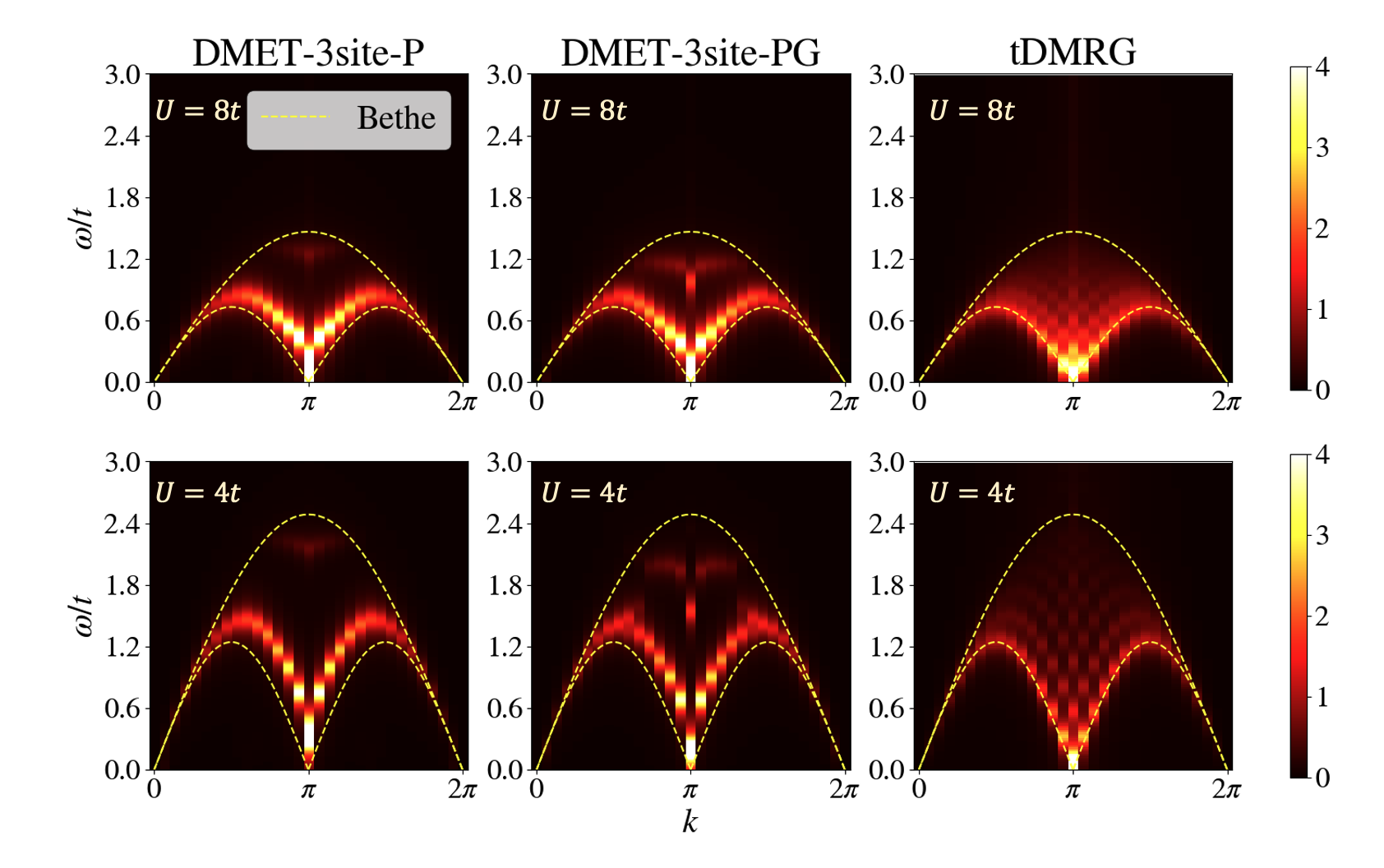}
    \caption{Spin dynamical structure factor $S(k, \omega)$ of the 30-site 1D Hubbard model from DMET (two excitation bases, $P$, $PG$, see text) and benchmark tDMRG results. Bethe ansatz results for the boundaries of the two-spinon excitations (in the TDL) are shown as yellow dashed lines. }
    \label{fig:gf_Sz}
\end{figure}

We focus on the case where $\hat A, \hat B$ are the site operators $\hat a_\sigma, \hat S_z = \dfrac{1}{2} (\hat a^\dagger_\alpha \hat a_\alpha - \hat a^\dagger_\beta \hat a_\beta), \hat N = \sum\limits_\sigma \hat a^\dagger_\sigma \hat a_\sigma$, corresponding to the single-particle Green's function, spin dynamical structure factor, and charge dynamical structure factor respectively. We use 3-site patches and two types of excitation operator bases, denoted by $P$ (projector) and $PG$ (projector plus generalized excitations): {\color{black}(see Table~\ref{table:PG-operators} for the precise list).} The cutoff of the overlap eigenvalues is set to 0.01 unless specified, and we compare against results from  time-dependent DMRG (tDMRG) \cite{Daley2004-aj,tddmrg-ddmrg-ronca,tddmrg-ren}, with time-step  $\Delta t = 0.2$ and a total evolution time of $T = 300$ for all the Green's functions. {\color{black}We use a broadening of $\eta=0.1$.}

We start with the spin dynamical structure factor $S(k, \omega)$ (Fig. \ref{fig:gf_Sz}). The first band of excited states calculated in the previous section corresponds to the lower boundary of the two-spinon excitations, as seen in the Bethe ansatz solution and the  tDMRG benchmark spectra. The DMET results captures this important feature, including the $k$-dependence of the spectral intensity, with the largest errors occurring near the maxima in the lower boundary.
However, the DMET spectra do not capture the continuum nature of the spectral density (between the two Bethe ansatz boundaries) seen in the tDMRG benchmark, due to the difficulty of representing higher-energy two-spinon excitations {\color{black}which go beyond the single-mode magnon ansatz}. We observe that if we increase the excitation basis size from basis $P$ to basis $PG$ we capture more intensity in the continuum region, generated by the additional spin-flips available in basis $PG$.  Additionally, similarly to the excitation energy, the shape of the two-spinon band improves when increasing the patch size, as shown in Figs. \ref{fig:spectra-Sz-2site} and \ref{fig:spectra-N-2site} in the Supplementary Information, which uses smaller two-site patches.

\begin{figure}[htbp]
    \centering
    \includegraphics[width=0.50\textwidth]{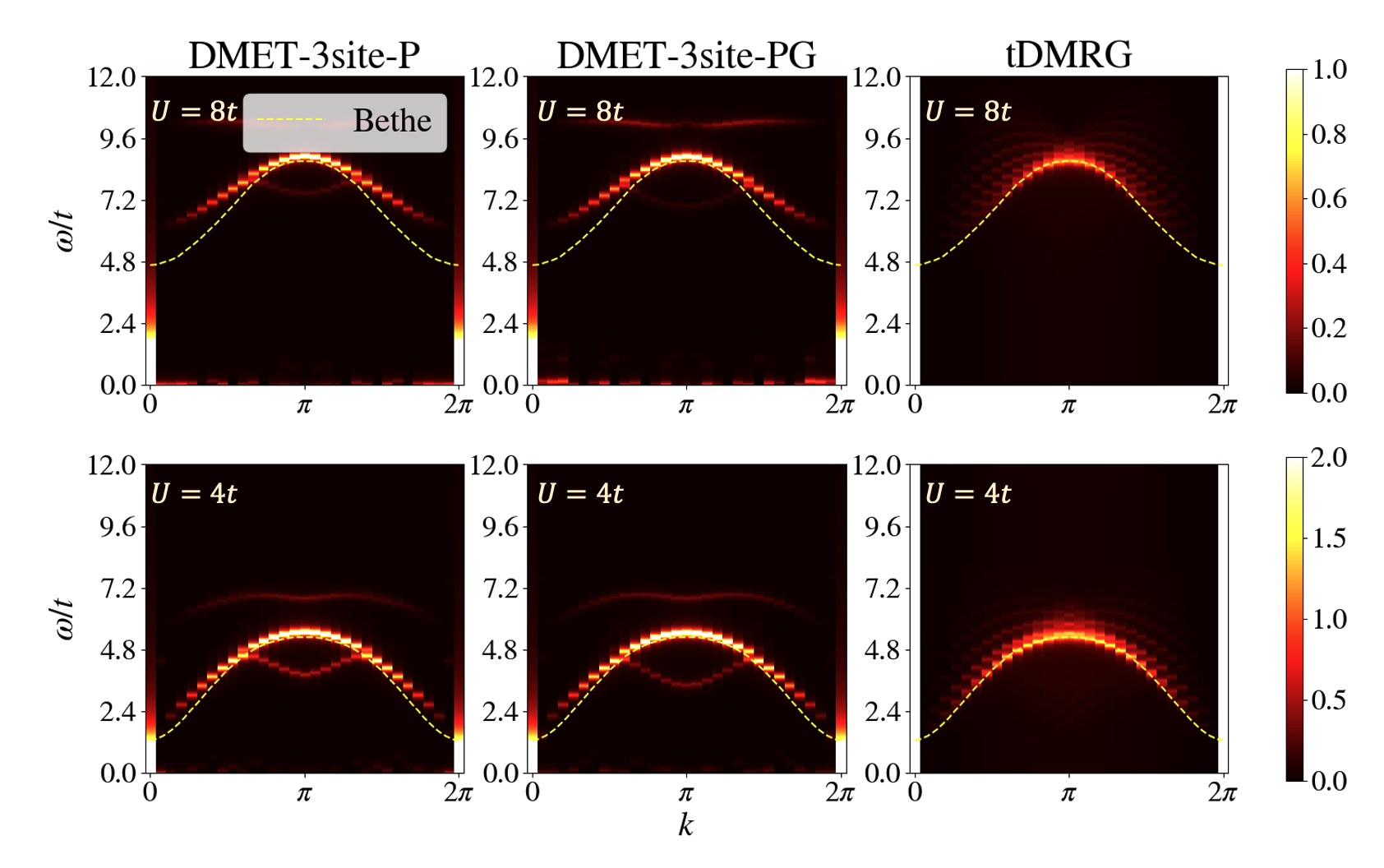}
    \caption{Charge dynamical structure factor $N(k, \omega)$ of 30-site 1D Hubbard model from DMET (excitation bases $P$ and $PG$, see text) and benchmark tDMRG results. Bethe ansatz results of the lower boundary of the band are shown by yellow dashed lines. The large intensity of $N(k, \omega)$ at $k=0, 2\pi$ comes from the ground-state contribution.}
    \label{fig:gf_N}
\end{figure}

\begin{figure}[htbp]
    \centering
    \includegraphics[width=0.50\textwidth]{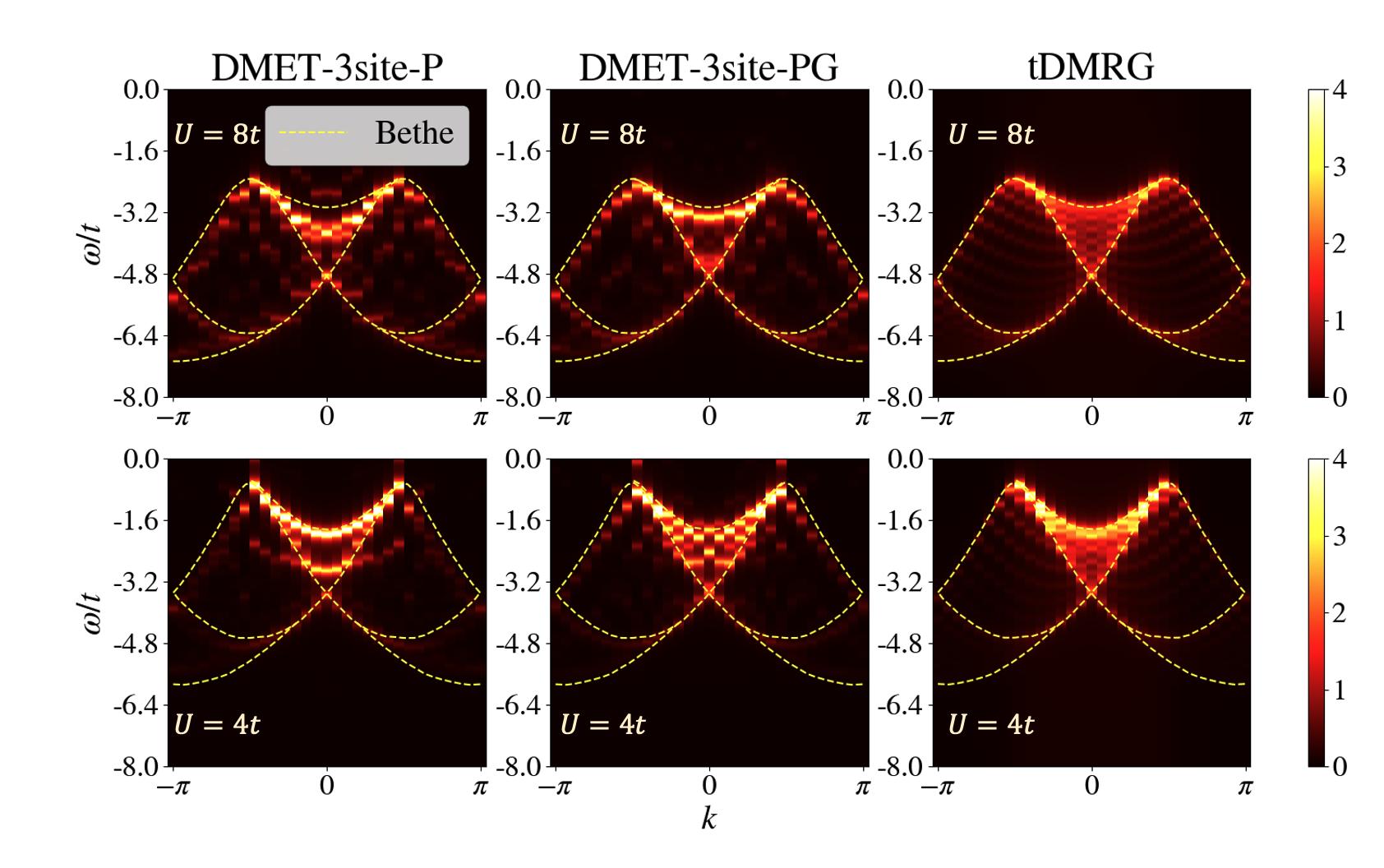}
    \caption{Spectral function $A(k, \omega)$ of 30-site 1D Hubbard model from DMET (excitation bases $P$, $PG$, see text) and benchmark tDMRG results. Boundaries of the spinon-holon dispersion from the Bethe Ansatz are shown by yellow dashed lines.}
    \label{fig:gf_d}
\end{figure}

The charge dynamical structure factor $N(k, \omega)$ is also qualitatively reproduced by our approximate DMET  spectra in Fig.~\ref{fig:gf_N}, although there is a significant overestimation of the excitation energy away from $k=\pi$. (Note that the large intensity at $k=0, 2\pi$ comes from the ground-state contribution to $N(k, \omega)$). Close to $k = \pi$ we also see some artifacts, with spectral weight at specific $k$ points at energies that lie below the benchmark lowest excitation energy for that $k$ point.  The single-particle spectral function $A(k,\omega)$ is obtained with similar quality
  (with only the lower branch depicted in Fig. \ref{fig:gf_d}, due to the particle-hole symmetry of the one-dimensional half-filled Hubbard model). Similarly to the tDMRG benchmark, the approximate DMET spectral function displays spinon and holon branches, and the boundary of the spinon-holon continuum is very close to that from the Bethe Ansatz.

\begin{figure*}[htbp]
        \centering
    \includegraphics[width=0.95\textwidth]{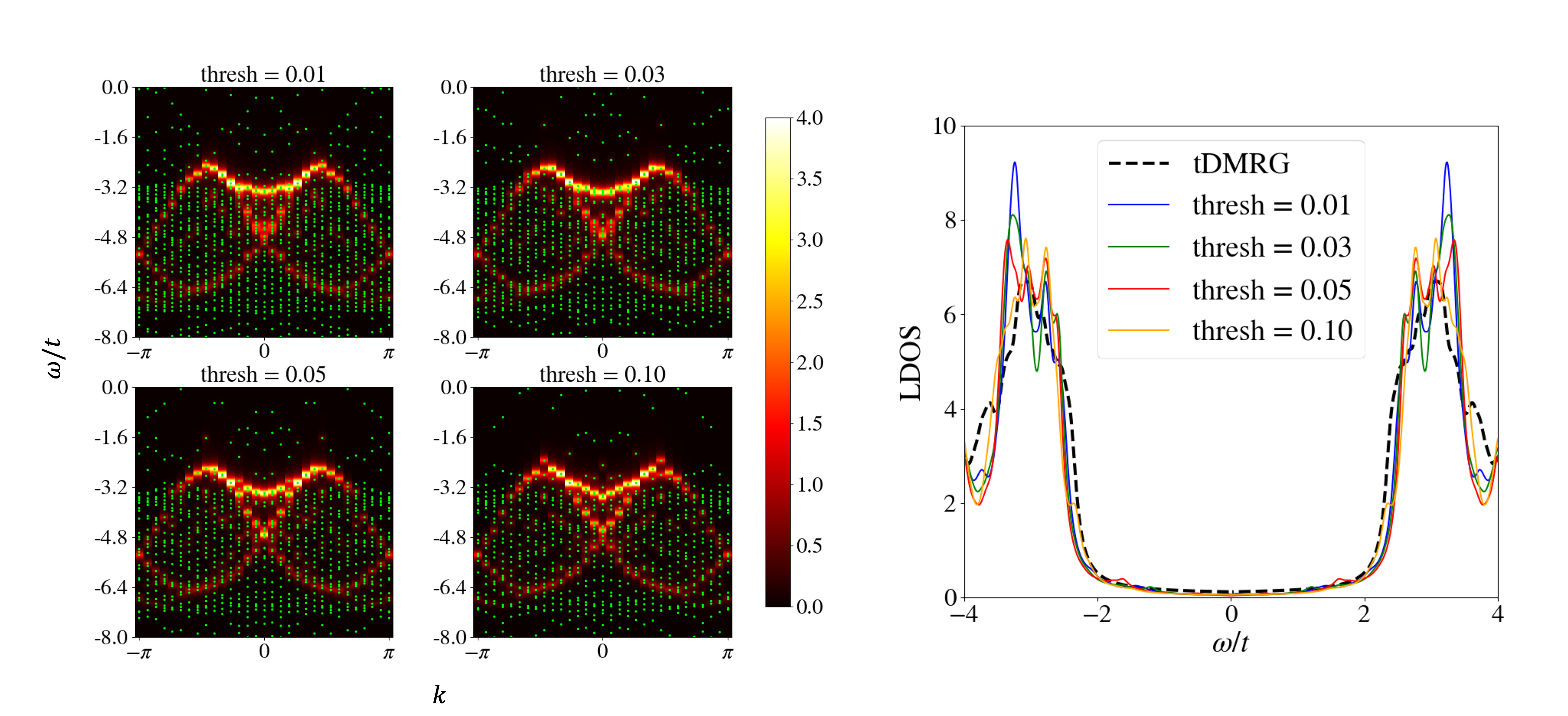}
    \caption{Spectral function (left) and local density of states (right) of the 1D Hubbard model $L=30, U=8$ with different overlap matrix eigenvalue thresholds. 3-site $PG$ ansatz is used for the DMET results. The k-space dispersion of the energy eigenvalues is shown by light green points on the spectral function plots.}
    \label{fig:spectra_thrd}
\end{figure*}

Fig. \ref{fig:spectra_thrd} shows the single-particle spectral function and energy eigenvalue dispersion computed from DMET using 
different overlap matrix eigenvalue thresholds. Eigenvalues above $\omega = -2.5t$ should be regarded as unphysical as they lie outside the exact upper boundary of the spinon-holon excitations from the Bethe ansatz. These unphysical excitations are sensitive to the overlap cutoff. Although present, they contribute only a small amount to the spectral weight and qualitatively resemble the tDMRG result (although they are visible in the unphysical weight outside of the Bethe ansatz boundaries of the spinon-holon dispersion in Fig.~\ref{fig:gf_d}). Further discussion of the unphysical states and the choice of overlap matrix eigenvalue thresholds is included in Section \ref{sec_si:thrd} of the Supplementary Information.

\begin{figure}
    \centering
    \includegraphics[width=0.45\textwidth]{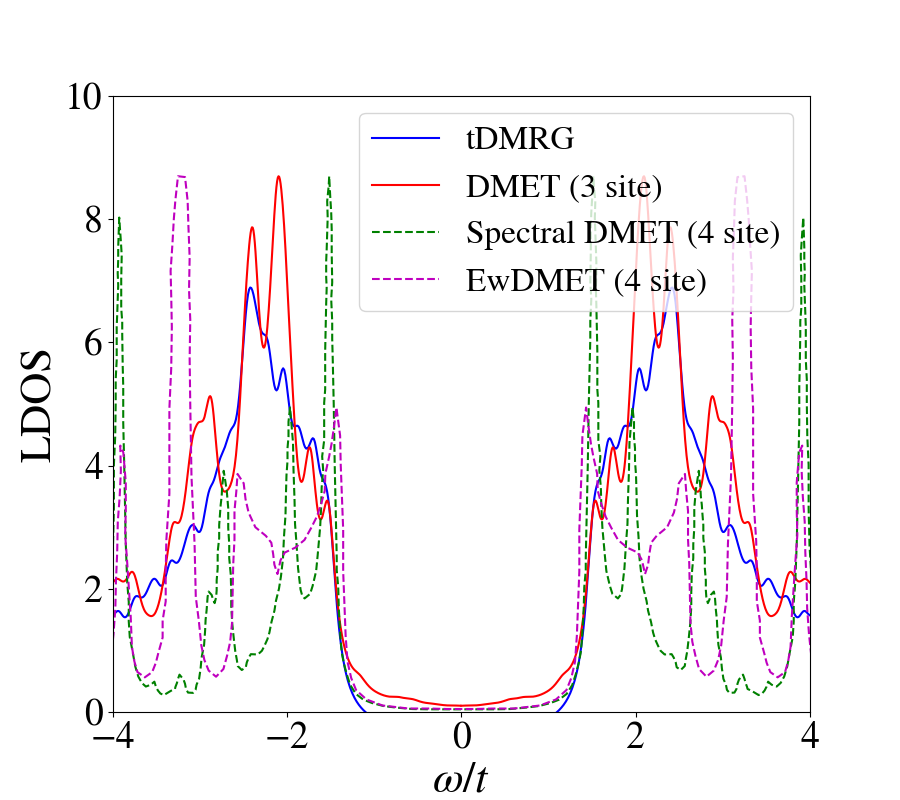}
    \caption{Comparison between the current DMET method (three-site patch, $PG$), the DMRG benchmark and other DMET-based approaches to compute the local density of states for the 1D Hubbard model, $U=6$. Results shown from spectral DMET (Ref. \cite{Booth2015-bl}) and energy-weighted DMET (EwDMET, Ref. \cite{Sriluckshmy2021-dz,Fertitta2019-rr}) extracted and redrawn via WebPlotDigitizer \cite{WebPlotDigitizer}.  tDMRG and DMET show the absolute LDOS value, {\color{black}{while the highest peaks of the spectral DMET and EwDMET results are scaled to the same height as the highest peak from the DMET data.}}  A broadening factor of $\eta = 0.1$ is used for all methods.} 
    \label{fig:gf_compare}
\end{figure}

Finally, we briefly compare the single-particle local density of states computed using our current scheme and those obtained from the spectral DMET \cite{Booth2015-bl} and the energy-weighted DMET (EwDMET) \cite{Sriluckshmy2021-dz,Fertitta2019-rr} schemes as shown in Fig. \ref{fig:gf_compare}. All the methods predict a similar spectral gap, however, away from the gap we find that the current approach yields a LDOS that is closer to the tDMRG benchmark. {We can calculate the similarity of LDOS $g$ and $f$ as
\begin{equation}
    S(f, g) = \sqrt{\int_{\omega_0}^{\omega_1} (f(\omega) - g(\omega))^2 \mathrm d \omega}
\end{equation}
where the smaller $S(f,g)$ is, the more similar the two spectra are. Taking $f$ to be the tDMRG reference and $g$ to be from the other methods, and $\omega_0 = -4t, \omega_1 = 0$, we obtain 1.7 for our method, 5.0 for spectral DMET and 4.5 for EwDMET, which demonstrates the closeness we achieve to the reference.}

\section{\label{sec:conclusion} Conclusions and Outlook}

We have described an approach to computing excited states and dynamical quantities within the density matrix embedding theory. The method is based on using a set of DMET impurities to approximate the matrix elements of the Hamiltonian within an excitation ansatz that captures collective excitations of the system. This allows us to go beyond earlier efforts based on embedding theories which focused on only the local spectral quantities. In tests on the one-dimensional Hubbard model, we find that we can capture many of the features of the excitations and spectral functions when compared to benchmark exact data.

Our current work,  however, also leaves considerable room for improvement. In particular (1) we have only used a ``single mode'' like excitation ansatz which does not allow us to describe all the low-lying excitations of the system, and (2) the DMET approximation to the matrix elements still leads to artifacts in the excitation space which we only partially succeed in removing. Improvements in these two directions will be important for the successful application of these methods to more challenging problems.

\begin{acknowledgments}
This project is sponsored by the US Department of Energy, Office of Science, Basic Energy Sciences, through Award No. DE-SC0018140. We thank Zhihao Cui and Shunyue Yuan for the development of the libdmet2 package that the excited-state code \cite{DMETExcitationPreview} is inspired by. We thank Runze Chi and Yuhang Ai for useful discussions.
\end{acknowledgments}

\section*{Author Contribution Statements}

\begin{itemize}
    \item Work conceptualization: G.K.C.(lead), C.L. (lead), S.L. (supporting)
    \item Code implementation: C.L.(excitation energies), S.L.(generalized excitation operators, Green's function and code refactoring), H.Z. (support for DMRG code)
    \item Validation: S.L.(equal), C.L.(equal)
    \item Data generation: S.L.(lead), C.L.(supporting)
    \item Writing - original draft: S.L.(lead), C.L.(supporting)
    \item Writing - review \& editing: S.L.(lead), G.K.C.(lead), C.L.(equal), H.Z.(equal)
\end{itemize}

\appendix

\section{\label{append:democratic-partitioning} Practical Estimation of Hamiltonian matrices by Democratic Partitioning}

We evaluate the transition density matrices with Eqs. \ref{eq:trans-1rdm} and \ref{eq:trans-2rdm} to illustrate the principle of democratic partitioning. In practice, we firstly use the partitioning rules on the lattice Hamiltonian, then the resulting operators are contracted by the excitation bases to generate the effective Hamiltonian and overlap matrix in Eq. \ref{eq:commutator}.

Consider the following Hamiltonian: 
\begin{equation}
    \hat H = \hat H_1 + \hat H_2 = \sum\limits_{pq} t_{pq} \hat a^\dagger_p \hat a_q + \dfrac{1}{2} \sum\limits_{pqrs} g_{pqrs} \hat a^\dagger_p \hat a_r^\dagger \hat a_s \hat a_q
\end{equation}
where $p,q,r,s$ are spin-orbital indices for sites. We denote the excitation basis as $| \Psi_{I} \rangle$. Looping around the choice of impurity sites, the transition Hamiltonian matrix element $\langle \Psi_{I} | \hat H | \Psi_{J} \rangle$ is partitioned between the impurities as
\begin{equation}
    \langle \Psi_{I} | \hat H | \Psi_{J} \rangle = \langle \Psi_{I} | \sum\limits_{x} \hat H_x | \Psi_{J} \rangle 
\end{equation}
and by using the DMET wave-function $| \Psi_{I}^x\rangle$ associated with the impurity $x$, we obtain the following approximation using democratic partitioning:
\begin{equation}
    \langle \Psi_{I} | \hat H | \Psi_{J}\rangle  = \sum\limits_{x} \langle \Psi_{I}^x | \hat H_x | \Psi_{J}^x \rangle = \sum\limits_x E_x
    \label{eq:approx-partition}
\end{equation}
where the partitioned Hamiltonian for each impurity $x$ is defined as in the ground-state partitioning, i.e.
\begin{equation}
\begin{aligned}
        \hat H^x &= \hat H_1^x + \hat H_2^x \\
        \hat H_1^x &=  \sum\limits_{p\in x} (\dfrac{1}{2} \sum\limits_{q} t_{pq} \hat a_p^\dagger \hat a_q + t_{qp} \hat a^\dagger_q \hat a_p) \equiv \sum\limits_{p\in A_x, q} \hat E_{pq} \\
        \hat H_2^x &= \sum\limits_{p\in A_x} \sum\limits_{qrs} \dfrac{1}{8} (g_{pqrs} \hat a^\dagger_p \hat a^\dagger_r \hat a_s \hat a_q + g_{qprs} \hat a^\dagger_q \hat a^\dagger_r \hat a_s \hat a_p \\
        &+ g_{qrps} \hat a^\dagger_q \hat a^\dagger_p \hat a_s \hat a_r + g_{qrsp} \hat a^\dagger_q \hat a^\dagger_s \hat a_p \hat a_r) \equiv \sum\limits_{p\in A_x, qrs} \hat e_{pqrs}
\end{aligned}
\label{eq:partitioned-hamil}
\end{equation}

The indices $p, q, r, s$ are originally defined in the site basis. After an orbital rotation to the DMET space of impurity, bath and unentangled orbitals (see Ref.~\cite{dmet-practical-guide-wouters} for additional details) for impurity $x$, denoted as $A_x$, $B_x$ and $C_x$ respectively, we derive the transition matrices $\langle \Psi_{I}^x | \hat E_{pq} | \Psi_{J}^x \rangle$ and $\langle \Psi_{I}^x | \hat e_{pqrs} | \Psi_{J}^x \rangle$:
\begin{equation}
    \langle \Psi_{I}^x | \hat E_{pq} | \Psi_{J}^x \rangle = \begin{cases}
        \dfrac{1}{2} (t_{pq} D_{pq} + t_{qp} D_{qp})&, q \in A_x + B_x; \\
        0&, q \in C_x.
    \end{cases}
\end{equation}

\begin{widetext}
\begin{equation}
    \begin{aligned}
        \langle \Psi_{I}^x | \hat e_{pqrs} | \Psi_{J}^x \rangle = \begin{cases}
            \frac{1}{8} (g_{pqrs} \Gamma_{pqrs} + g_{qprs} \Gamma_{qprs} + g_{qrps} \Gamma_{qrps} + g_{qrsp} \Gamma_{qrsp} ) &,\, q,r,s \in A_x + B_x ; \\
            \frac{1}{8} (g_{pqrs} D_{sr} D_{qp} + g_{qprs} D_{pq} D_{sr} - g_{qrsp} D_{pq} D_{rs}) &, \, q \in A_x + B_x;\; r,s \in C_x; \\
            \frac{1}{8} (-g_{pqrs} D_{qr} D_{sp} + g_{qrps} D_{rq} D_{sp} + g_{qrsp} D_{rq} D_{ps}) &,\, s \in A_x + B_x ,\, q, r \in C_x; \\
            -\frac{1}{8}(g_{pqrs}D_{sq}D_{pr} + g_{qrps} D_{sq}D_{rp})&,\, r \in A_x, B_x ;\; q, s \in C_x.
        \end{cases}
    \end{aligned}
\end{equation}
\end{widetext}
where the reduced density matrices $\Gamma_{pqrs}, D_{pq}$ are defined by the embedding-space or core-space wave functions:
\begin{equation}
    \begin{aligned}
    \Gamma_{pqrs} &= \langle \Psi_{I}^{\mathrm{emb},x} | \hat a^\dagger_p \hat a^\dagger_r \hat a_s \hat a_q | \Psi_{J}^{\mathrm{emb},x} \rangle \\
    D_{pq} &= \begin{cases}
        \langle \Psi_{I}^{\mathrm{emb},x} | \hat a^\dagger_q \hat a_p | \Psi_{J}^{\mathrm{emb},x} \rangle &,\, p, q \in A_x + B_x \\
        \langle \gamma | \hat a^\dagger_q \hat a_p | \gamma \rangle &,\, p, q \in C_x.
    \end{cases}
    \end{aligned}
    \label{eq:emb-rdm12-notation}
\end{equation}
where $|\gamma \rangle$ is the Slater determinant of unentangled core orbitals.  Eq. \ref{eq:emb-rdm12-notation} gives a compact formula to evaluate the transition Hamiltonian matrix for each impurity $E_x$
\begin{equation}
    \begin{aligned}
        E_x = \sum\limits_{p}^{A_x} (\dfrac{1}{2} \sum\limits_{q}^{A_x + B_x} f_{pq} D_{qp} + \dfrac{1}{4} \sum\limits_{qrs}^{A_x + B_x} g_{pqrs} \Gamma_{pqrs}
    \\ + \dfrac{1}{2} \sum\limits_{q}^{A_x + B_x} f_{qp} D_{pq} + \dfrac{1}{4} \sum\limits_{qrs}^{A_x + B_x} g_{qprs} \Gamma_{qprs})
    \end{aligned}
    \label{eq:Ex-by-rdm}
\end{equation}
where the Fock matrix $f_{pq}$ is defined as
\begin{equation}
    f_{pq} + \sum\limits_{rs}^{C_x} \dfrac{1}{2} (g_{pqrs} - g_{psrq}) D_{rs}
\end{equation}

 Eq. \ref{eq:Ex-by-rdm} can be thought of as obtained from an  effective second-quantized Hamiltonian $\tilde H_x$ ($E_x = \langle \Psi_I^{\mathrm{emb}, x} | \tilde H_x | \Psi_J^{\mathrm{emb, x}} \rangle $) in the embedding space $A_x+B_x$:
\begin{equation}
    \tilde H_x = \dfrac{1}{2} \sum\limits_{pq}^{A_x + B_x} f_{pq} w_{pq} \hat a^\dagger_p \hat a_q + \dfrac{1}{8} \sum\limits_{pqrs}^{A_x + B_x} g_{pqrs} w_{pqrs} \hat a^\dagger_p \hat a^\dagger_r \hat a_s \hat a_q
\end{equation}
where the additional coefficients $w_{pq}$ and $w_{pqrs}$ denote the number of indices in the impurity $A_x$. For instance, $w_{pqrs} = 3$ when indices $p, q, r$ are in the impurity and $s$ is in the bath.

\section{\label{append:PPM-projectors} Specification of the generalized excitation bases}

In section \ref{subsec:gf}, we use a generalized excitation basis (denoted as PG in that context) truncated from the full general excitation basis. For neutral excitations, the excitation basis contains single-site  projectors $\hat P_{s_i}$ as well as the spin-flip operators $\hat s_{\pm}$.  Destruction operators $\hat a_\alpha, \hat a_\beta$ are further included in addition to $\hat P_{s_i}$, $\hat s_\pm$, when computing the single-particle Green's function.

\begin{table}[htbp]
    \centering
    \begin{tabular}{cccc}
    \hline
    $N$ & $S_z$ & Operators & Count \\
    \hline
    0 & 0 & $\hat P \otimes \hat P \otimes \hat P$ & 64 \\
    \hline \\ 
    -1 & -1/2 & $\hat P \otimes \hat P \otimes \hat A_\alpha,\hat P \otimes \hat A_\alpha \otimes  \hat P $ & 96 \\
    & & $\hat A_\alpha \otimes \hat P \otimes \hat P$ & \\
    \hline
    
    \end{tabular}
    \caption{3-site $P$ operator set for calculating neutral excitations and charged excitations. $\hat P \in \{\hat n_\alpha \hat n_\beta, \hat n_\alpha (1 - \hat n_\beta), (1 - \hat n_\alpha) \hat n_\beta, (1 - \hat n_\alpha)(1 - \hat n_\beta)\}$, $\hat A_\alpha \in \{\hat a_\alpha \hat n_\beta, \hat a_\alpha (1 - \hat n_\beta)\}$}.
    \label{table:P-operators}
\end{table}

\begin{table}[htbp]
    \centering
    \begin{tabular}{ccccc}
    \hline
    $N$ & $S_z$ & Operators & Count & Total \\
    \hline
    0 & 0 & $\hat P \otimes \hat P \otimes \hat P$ & 64 & 88 \\
    & & $\hat P \otimes \hat s_+ \otimes \hat s_-, \hat P \otimes \hat s_- \otimes \hat s_+$ & 24 \\
    & & $\hat s_+ \otimes \hat P \otimes \hat s_-, \hat s_-  \otimes  \hat P \otimes \hat s_+$ &  \\
    & & $\hat s_+ \otimes \hat s_- \otimes \hat P,  \hat s_- \otimes \hat s_+ \otimes \hat P$ \\ \hline \\ 
    -1 & -1/2 & $\hat P \otimes \hat P \otimes \hat A_\alpha,\hat P \otimes \hat A_\alpha \otimes  \hat P $ & 96 & 156 \\
    & & $\hat A_\alpha \otimes \hat P \otimes \hat P$ & & \\
    & & $\hat P \otimes \hat s_- \otimes \hat A_\beta , \hat s_- \otimes \hat A_\beta \otimes \hat P$ & 48 \\
    & & $\hat A_\beta \otimes \hat P \otimes \hat s_-, \hat P \otimes   \hat A_\beta \otimes \hat s_-$ \\
    & & $ \hat s_- \otimes   \hat P \otimes \hat A_\beta, \hat A_\beta \otimes \hat s_- \otimes \hat P $ \\
    & & $\hat s_+ \otimes \hat s_- \otimes \hat A_\alpha, \hat s_+ \hat A_\alpha \otimes \hat s_-$ & 12 \\
    & & $\hat A_\alpha \otimes \hat s_+ \otimes \hat s_-, \hat s_- \otimes \hat s_+ \otimes \hat A_\alpha$ \\
    & & $\hat s_- \hat A_\alpha \otimes \hat s_+, \hat A_\alpha \otimes \hat s_- \otimes \hat s_+$ \\
    \hline
    
    \end{tabular}
    \caption{3-site $PG$ operator set for calculating neutral excitations ($ N =  S_z = 0$) and charged excitations ($ N = 1,  S_z = -\dfrac{1}{2}$). $\hat A_\beta \in \{ \hat a_\beta \hat n_\alpha , \hat a_\beta (1 - \hat n_\alpha)\}$}.
    \label{table:PG-operators}
\end{table}

\bibliography{main}

\onecolumngrid

\renewcommand{\thesubsection}{S\arabic{subsection}}
\renewcommand{\thefigure}{S\arabic{figure}}
\renewcommand{\thetable}{S\arabic{table}}
\renewcommand{\theequation}{S\arabic{equation}}

\setcounter{section}{0}
\setcounter{figure}{0}
\setcounter{table}{0}
\setcounter{equation}{0}

\section*{Supplementary Information}

\subsection{\label{sec_si:thrd} Quantitative analysis of threshold}

\begin{figure}[htbp]
    \centering
    \includegraphics[width=0.6\linewidth]{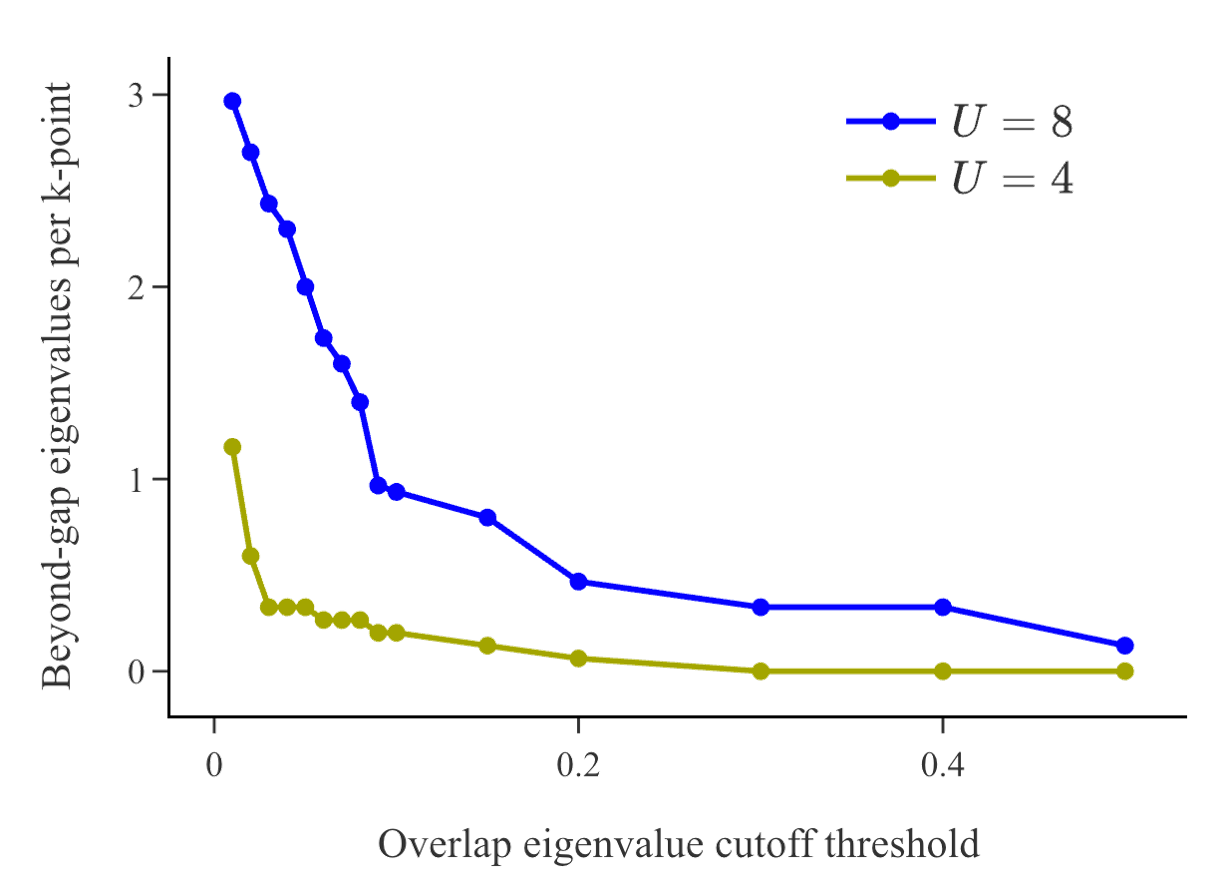}
    \caption{The number of effective Hamiltonian eigenvalues beyond the upper bound of $A(k, \omega)$ for different overlap eigenvalue cutoffs.}
    \label{fig:in-gap-number}
\end{figure}

\begin{figure}[htbp]
    \centering
    \includegraphics[width=0.75\linewidth]{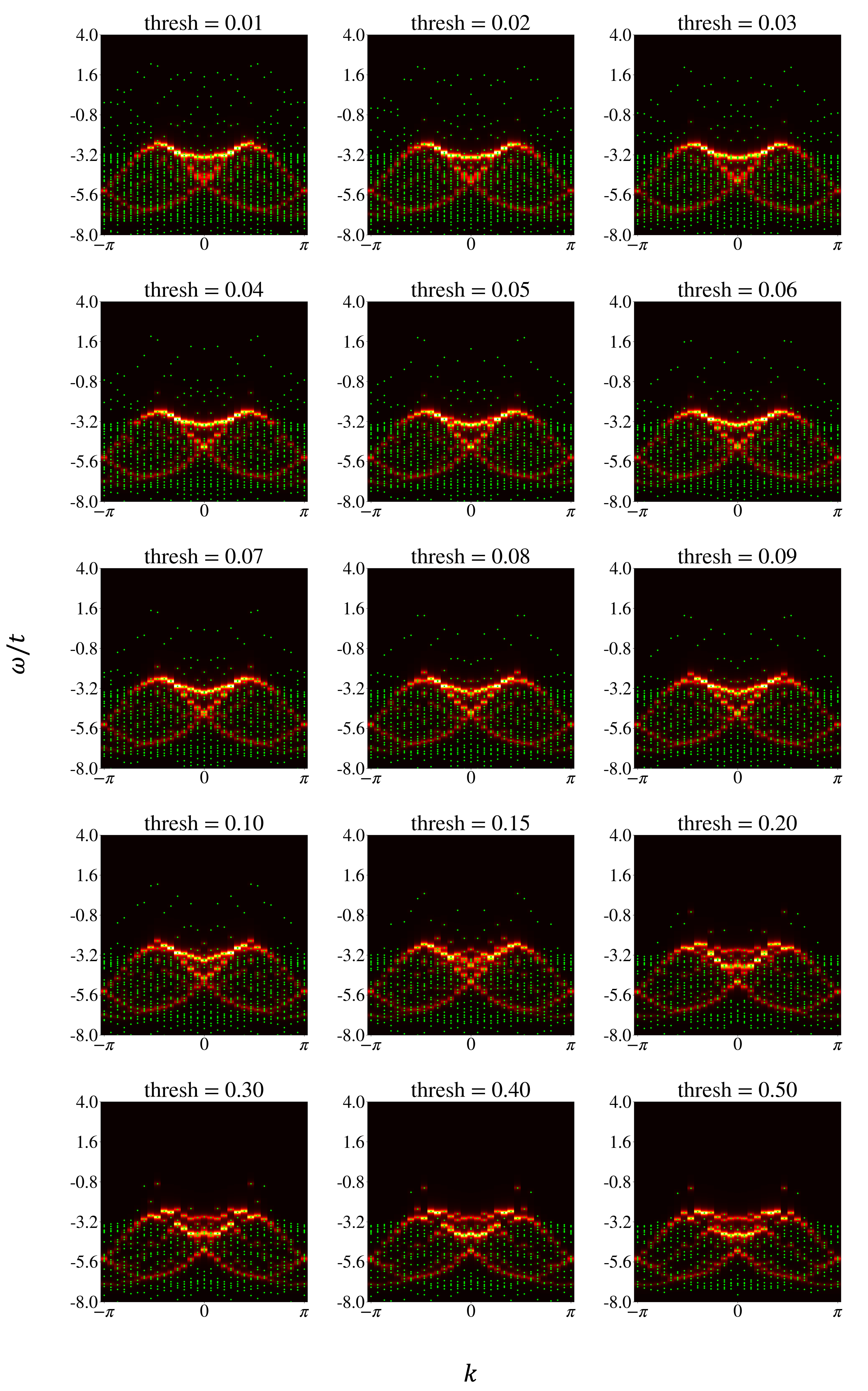}
    \caption{$A(k, \omega)$ of the Hubbard model $L=30, U=8$ with a larger range of overlap cutoff threshold ($0.01 \sim 0.50$).}
    \label{fig:larger-range-cutoff-test}
\end{figure}

\begin{figure}[htbp]
    \centering
    \includegraphics[width=0.75\linewidth]{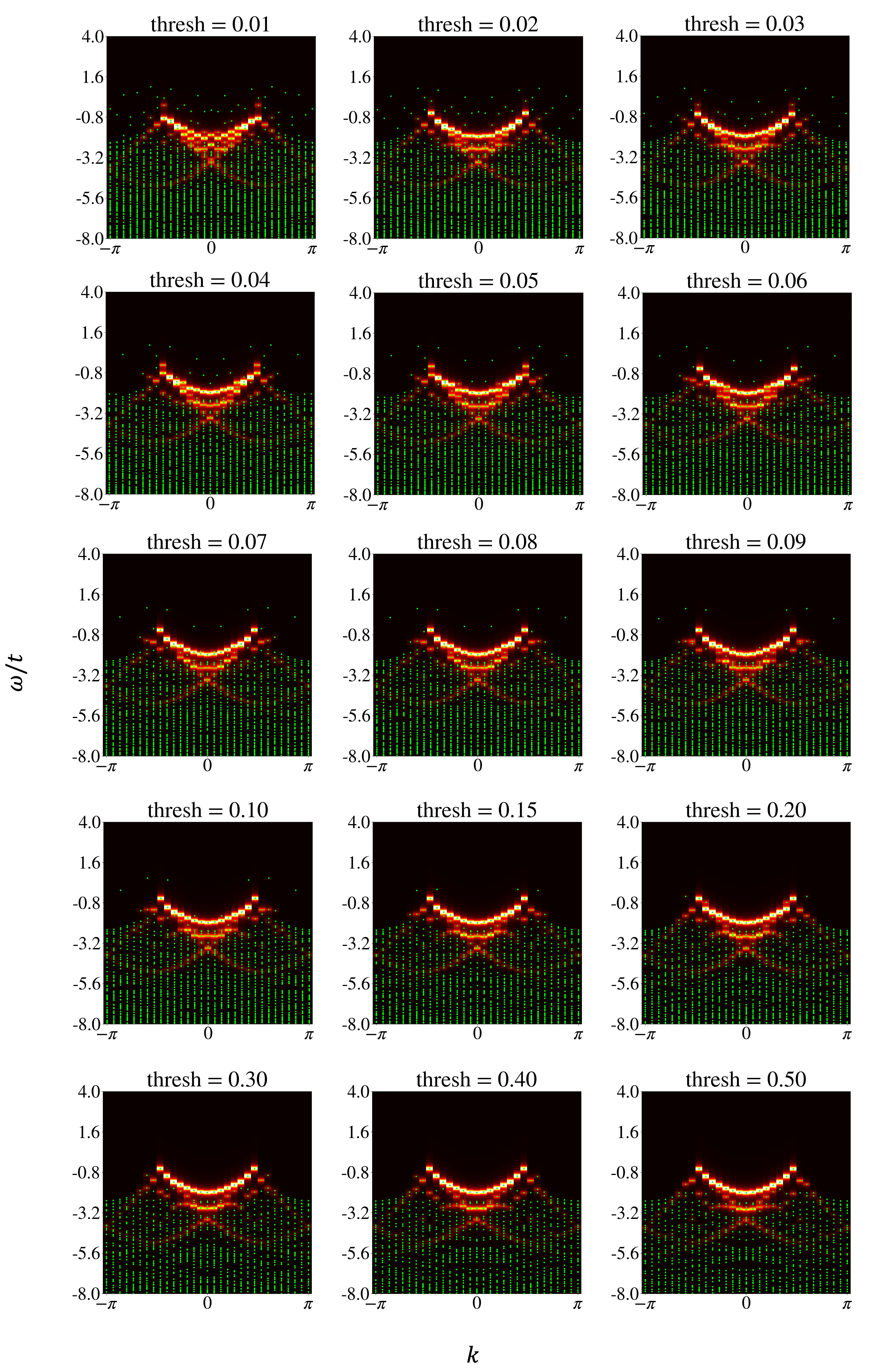}
    \caption{$A(k, \omega)$ of the Hubbard model $L=30, U=4$ with a larger range of overlap cutoff threshold ($0.01 \sim 0.50$).}
    \label{fig:larger-range-cutoff-test-u4}
\end{figure}

We test a broader range of overlap matrix eigenvalue cutoffs and observe its influence on the charged excited states and spectral function $A(k, \omega)$ in Fig. \ref{fig:larger-range-cutoff-test} and Fig. \ref{fig:larger-range-cutoff-test-u4}, and count the number of eigenvalues beyond the upper bound of $A(k, \omega)$ as a rough measurement of the quantity of unphysical eigenstates, shown in Fig. \ref{fig:in-gap-number}.

Both the $U=8$ and $U=4$ lines of Fig. \ref{fig:in-gap-number} show the same trend of a decrease in the number of unphysical eigenstates: The number of unphysical states drops rapidly before some threshold (0.1 for $U=8$ and 0.04 for $U=4$), then it slowly decreases after further increasing the threshold. The profile of the spectra slightly change before threshold $0.04$ for $U=4$, and then the spectral shape is stable, while for $U=8$ shown in Fig. \ref{fig:larger-range-cutoff-test}, the shape of the spectra develops artifacts at larger thresholds. Nonetheless, the choice of overlap cutoff threshold remains empirical in practice.


\subsection{\label{sec_si:patch-size}Dependence on patch sizes of Green's functions}

\begin{figure}[htbp]
    \centering
    \includegraphics[width=0.85\linewidth]{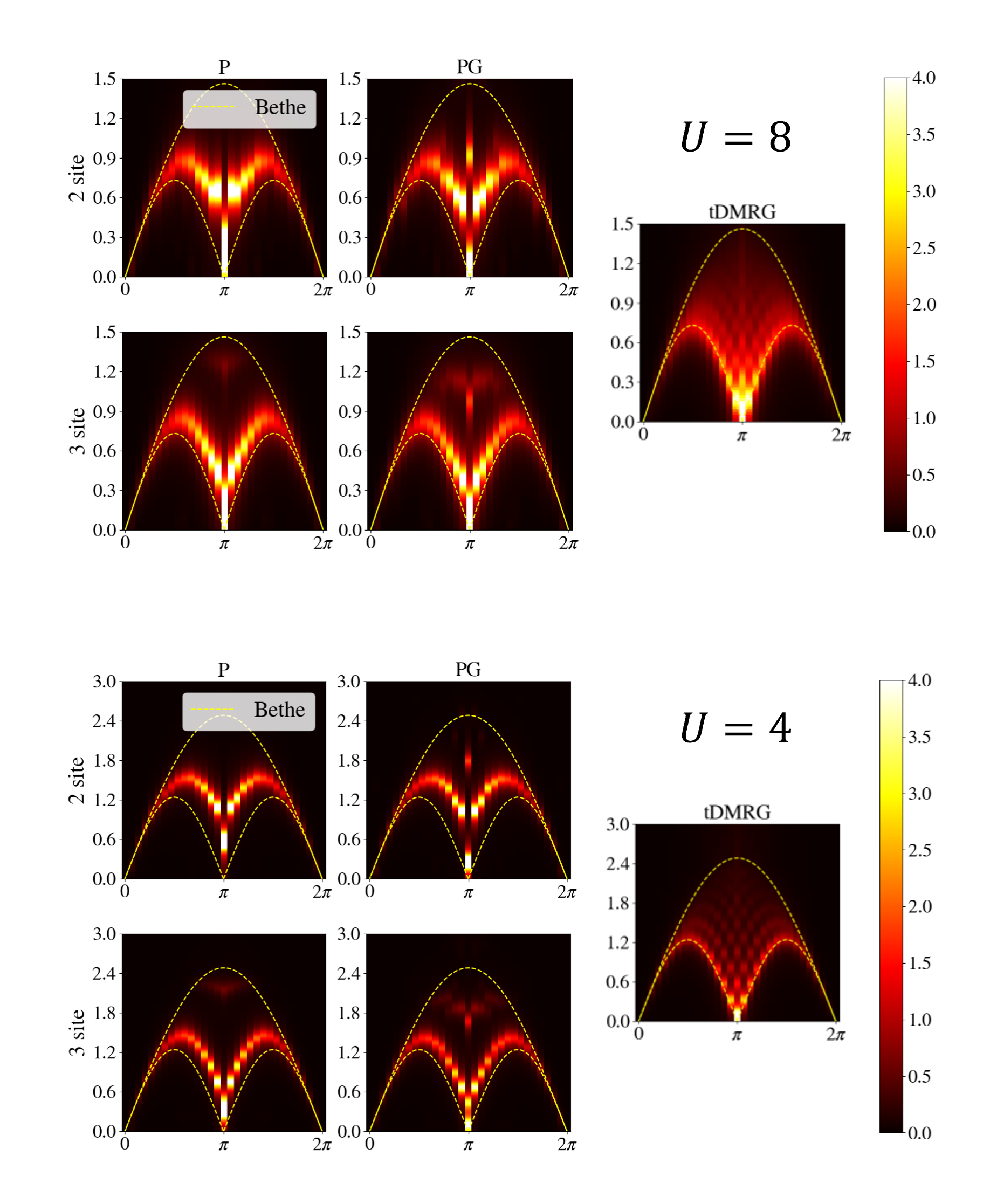}
    \caption{Spin dynamical structure factor $S(k, \omega)$ of the 30-site 1D Hubbard model from DMET (two-site and three-site patches), benchmark tDMRG results and Bethe Ansatz.}
    \label{fig:spectra-Sz-2site}
\end{figure}

\begin{figure}[htbp]
    \centering
    
    \includegraphics[width=0.85\linewidth]{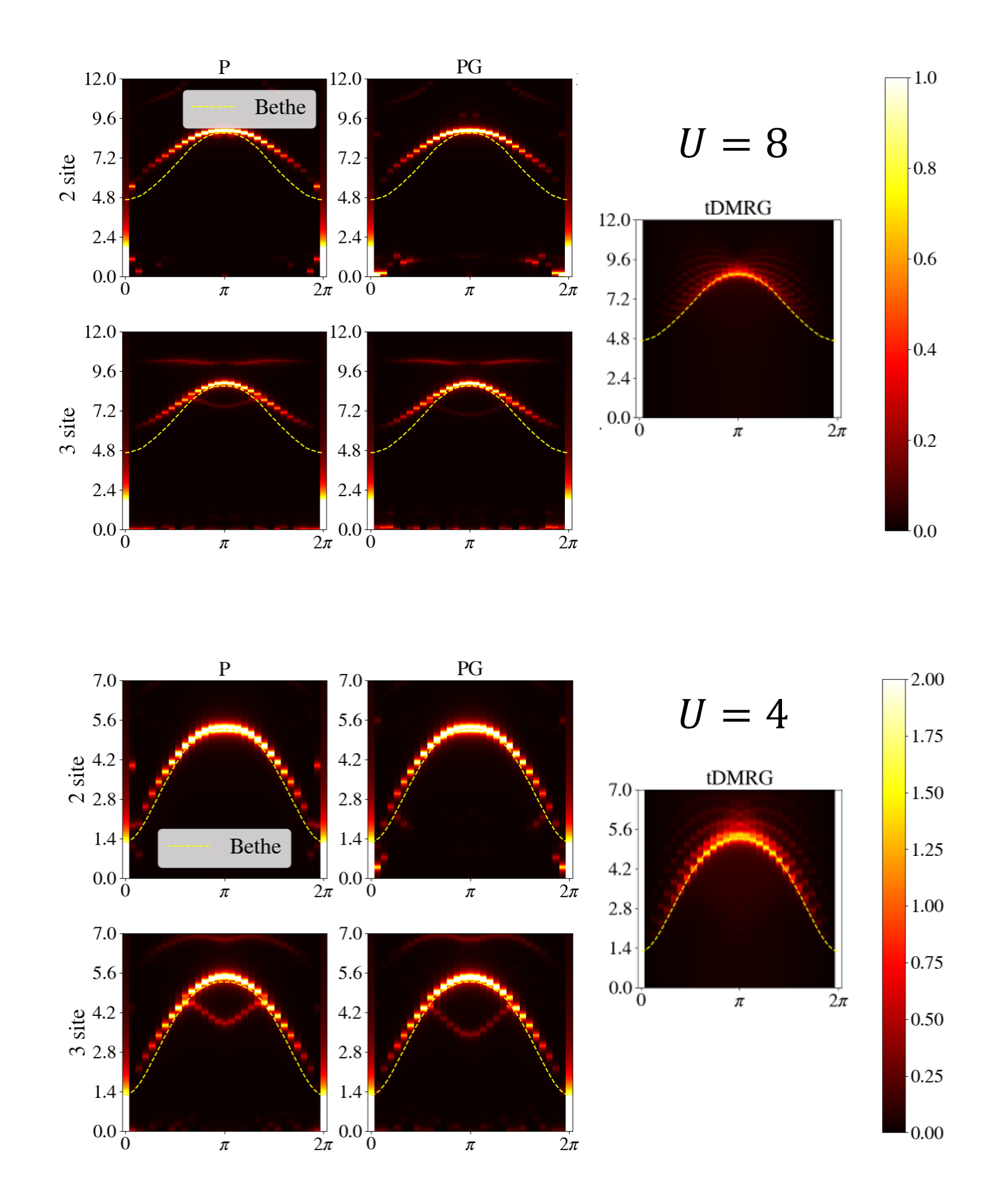}
    \caption{Charge dynamical structure factor $N(k, \omega)$ of the 30-site 1D Hubbard model from DMET (two-site and three-site patches), benchmark tDMRG results and Bethe Ansatz.}
    \label{fig:spectra-N-2site}
\end{figure}

We illustrate the dependence on patch size with additional computation using two-site patches in Figs. \ref{fig:spectra-Sz-2site} and \ref{fig:spectra-N-2site} using the excitation operator ansatz  $P$ and $PG$ listed in Table \ref{table:P-operators-2site} constructed in the same way as for three-site patches (Table \ref{table:P-operators} and \ref{table:PG-operators}. We clearly observe the improvement of the $S(k, \omega)$ structural factor from the two-site $P/PG1$ ansatz to the three-site $P/PG$ ansatz, especially at $\dfrac{\pi}{2} < k < \pi$ where the dispersion of the three-site patch is closer to that of the Bethe Ansatz and tDMRG. However the improvement is harder to see in the charge structure factor $N(k, \omega)$.

\begin{table}[htbp]
    \centering
    \begin{tabular}{ccc}
    \hline
    Type & Operators & Count \\
    \hline
    P  & $\hat P \otimes \hat P$ & 16 \\
    \hline \\ 
    PG1 & $\hat P \otimes \hat P, \hat s_+ \otimes \hat s_-, \hat s_- \otimes \hat s_+$ & 18 \\
    \hline
    \end{tabular}
    \caption{2-site operator sites for calculating neutral excitations. $\hat P \in \{\hat n_\alpha \hat n_\beta, \hat n_\alpha (1 - \hat n_\beta), (1 - \hat n_\alpha) \hat n_\beta, (1 - \hat n_\alpha)(1 - \hat n_\beta)\}$.}
    \label{table:P-operators-2site}
\end{table}

\end{document}